# Proposed alteration of images of molecular orbitals obtained using a scanning tunnelling microscope as a probe of electron correlation

Dimitrios Toroz, Massimo Rontani*, Stefano Corni[†]

*CNR-NANO Research Center S3, Via Campi 213a, 41125 Modena, Italy*

Scanning tunneling spectroscopy (STS) allows to image single molecules decoupled from the supporting substrate. The obtained images are routinely interpreted as the square moduli of molecular orbitals, dressed by the mean-field electron-electron interaction. Here we demonstrate that the effect of electron correlation beyond mean field qualitatively alters the uncorrelated STS images. Our evidence is based on the ab-initio many-body calculation of STS images of planar molecules with metal centers. We find that many-body correlations alter significantly the image spectral weight close to the metal center of the molecules. This change is large enough to be accessed experimentally, surviving to molecule-substrate interactions.

PACS numbers: 74.55.+v, 31.15.vq, 82.37.Gk, 73.23.Hk

Scanning tunnelling spectroscopy (STS) visualizes electron states in both extended systems and nano-objects, such as quantum dots or molecules [1,2]. Whereas extended quantum states are insensitive to electron number fluctuations, an energy gap opens each time a new electron is injected by the STS tip into a nano-object. This gap originates from the interaction of the next incoming electron with the others already present in the system. Under this Coulomb blockade condition [3--5], STS maps the wave function modulus of the electron injected by the tip into the nano-object. The obtained image is routinely interpreted as the atomic-like [6--10] or molecular [11--15] orbital of the added electron, that experiences the mean field of the other electrons already populating the system. A fundamental question is whether features of the tunnelling map may appear due to electron-electron correlation beyond mean field [16--20].

In this Letter we demonstrate that the STS images of single planar molecules with metal centres predicted by ab initio many-body calculations differ qualitatively from their uncorrelated counterparts. We find in the STS maps resolved at the Fermi energy that correlation alters significantly the spectral weight around the metal atom. This change may be experimentally quantified by contrasting the altered STS images to those of substituted molecules unaffected by correlation that are used as benchmarks.

As a concrete example we propose two cousin complexes differing in the metal centre species, which is either copper or zinc. Whereas the copper-saloph exhibits STS images strongly distorted by correlation, the substituted zinc-saloph is hardly correlated. We predict that, at the many-body level, the images of the two positively ionized complexes are identical at large tip-molecule distance (small current through the tip) but differ at small distance (large current). At the uncorrelated level the trend is opposite, with the two images overlapping at closer tip-molecule distance but parting at larger distance. Therefore, the experimental validation of the many-body effect requires that: (i) the measured STS maps of copper- and zinc-saloph become alike as the tip-molecule distance is increased (ii) the STS tip resolution---to be estimated independently---is adequate.

Many-body phenomena such as Kondo effect [21--24] and Fermi edge singularity [25] have already been observed in molecules and quantum dots. Moreover, quasiparticle calculations have been used to predict the alignment of molecular levels with respect to the extended states of the supporting substrate [26--28]. Whereas the above effects are induced by various types of interactions between the molecule and either the substrate or the STS tip, here we focus on the intrinsic few-body behaviour of almost isolated molecules. A prerequisite is the electrical decoupling of the molecule from the conductive substrate, that may be achieved either by inserting a thin insulating overlayer [11--13,15] or by physisorbing the molecule [14].

Molecules with metal centers are targets of choice to investigate effects of correlation, as they sustain different electronic configurations of comparable energies [23,24,29]. We focus on planar complexes in order to simplify the analysis of STS maps, starting from a prototypical system, the copper-chloride ion ($CuCl_4^{2-}$) in square planar arrangement. This complex consists in a Cu atom at the center of a square with four Cl atoms at the corners (see inset in Fig. 1). This square symmetry is apparent from the contour plots of the molecular orbitals, computed in the standard framework of the Hartree-Fock (HF) mean-field theory after the preliminary optimization of the molecular geometry (left column of Fig. 1). Since the ground state is a spin doublet, the highest (singly) occupied molecular orbital (labelled SOMO-HF) is very similar to the lowest unoccupied one (LUMO-HF), with small discrepancies arising from the spin-unrestricted character of the calculation [30].

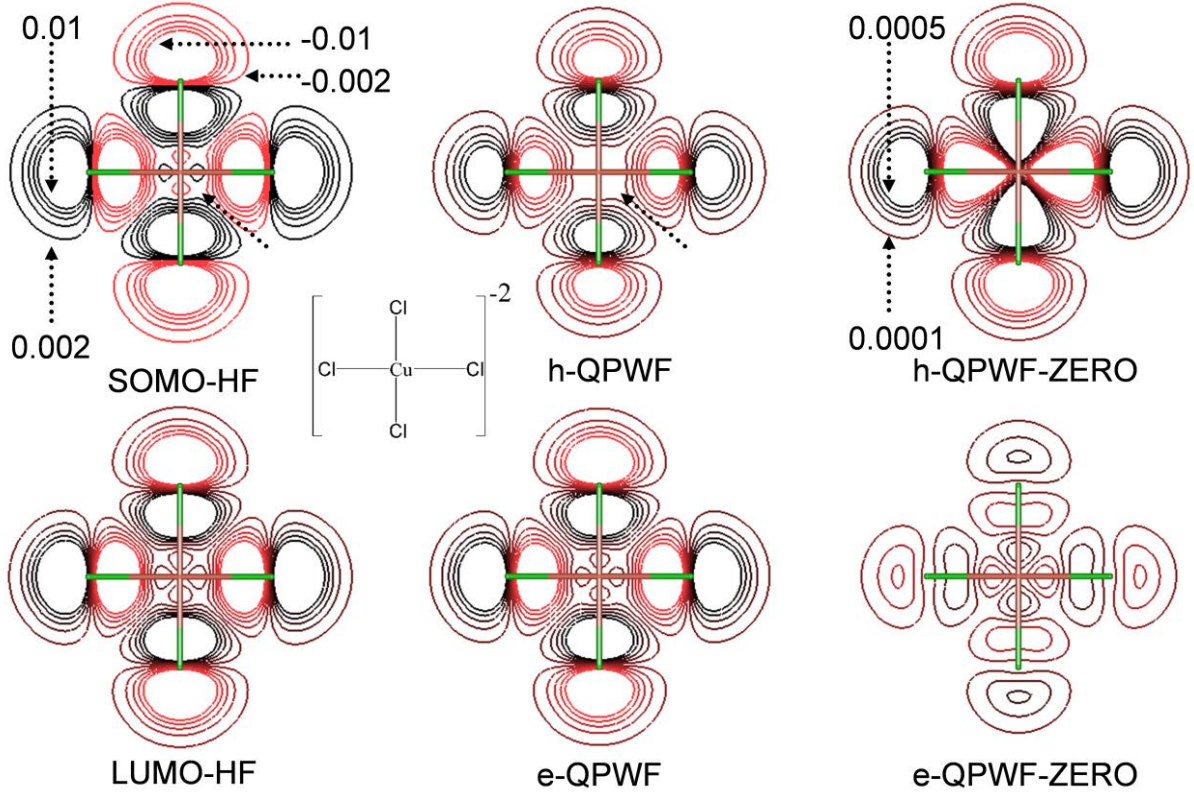

FIG. 1 (color online). Molecular orbitals and quasiparticle wave functions of copper chloride. Contour plots of Hartree-Fock frontier orbitals of copper chloride (SOMO-HF and LUMO-HF of the *N*-state, left column), quasiparticle wave functions (h-QPWF and e-QPWF, center column), and QPWFs omitting the contribution of frontier orbitals (h-QPWF-ZERO and e-QPWF-ZERO, right column). All contours are taken 2.0 Å over the molecular plane and parallel to it. Black (red) contours have positive (negative) values. The contour line maximum (minimum) value is $10^{-2}$ Å$^{-3/2}$ ($-10^{-2}$ Å$^{-3/2}$) for HOMO, LUMO, QPWFs, and $5 \cdot 10^{-4}$ Å$^{-3/2}$ ($-5 \cdot 10^{-4}$ Å$^{-3/2}$) for QPWF-ZEROs. Brown (green) segments of the molecule backbone indicate the Cu (Cl) sites. Inset: chemical structure of copper chloride.

To investigate how correlation affects HF theory, we introduce the quasiparticle wave function [16--19] $\varphi(\mathbf{r})$ of the electron added to the molecule (e-QPWF),

$$\varphi(\mathbf{r}) = \langle N+1 | \hat{\Psi}^+(\mathbf{r}) | N \rangle. \qquad (1)$$

The e-QPWF $\varphi(\mathbf{r})$ is the probability amplitude of finding an electron---added to the neutral ground state $|N\rangle$ by the field operator $\hat{\Psi}^+(\mathbf{r})$ at position $\mathbf{r}$---in the negatively ionized molecule ground state $|N+1\rangle$. In the absence of correlation, $|N\rangle$ and $|N+1\rangle$ are single Slater determinants and the e-QPWF is reduced to the LUMO-HF as a consequence of Koopmans' theorem. This observation endows LUMO-HF with a clear physical meaning, amenable to experimental measurement when electron correlation is weak. In the correlated case, the e-QPWF must be computed through (1) from the knowledge of both correlated ground states $|N\rangle$ and $|N+1\rangle$. Similar considerations hold for the hole QPWF (h-QPWF), $\varphi(\mathbf{r}) = \langle N-1|\hat{\Psi}(\mathbf{r})|N\rangle$.

In practice, to compute the e-QPWF we perform a restricted HF calculation on the $N + 1$ ground state, that is a closed-shell system for all considered molecules. We save and then employ the HF molecular orbitals, with no further changes, for the calculations of the $N + 1$ and $N$ molecular states by using a correlated method, either coupled cluster or configuration interaction [30], depending on the molecule size. For the h-QPWF, the procedure is the same but we use the $N$ -1 state instead of $N$ + 1. We perform all quantum mechanical calculations with Gaussian 09 suite of codes [31], using the Lanl2DZ basis set and effective core potentials (cf. Supplementary Discussion S2 for the dependence of the wave function tail on the basis set [33]). We compute QPWFs by means of an in-house parallel code.

We obtain the h-QPWF and e-QPWF of copper chloride (middle column of Fig. 1), forced in a planar geometry, through coupled cluster calculations (CCSD)---a highly accurate method to include correlation effects in molecular states [30]. Amplitude weight on the metallic centre is present for negative ionization (e-QPWF) but absent for positive ionization (h-QPWF): Indeed, a correlation hole forms in the h-QPWF, in the region indicated by the arrow, whereas significant amplitude is present in the same position in the uncorrelated orbital (SOMO-HF, top left panel).

The differences between the h-QPWF and the HF SOMO are caused by the interference among the orbitals $\phi_\alpha(\mathbf{r})$ of different electronic configurations that contribute to $\varphi(\mathbf{r})$, according to [19]

$$\varphi(\mathbf{r}) = \sum_{i,j} (C_j^{N-1})^* C_i^N \sum_\alpha \phi_\alpha(\mathbf{r}) \langle \Phi_j^{N-1} | \hat{c}_\alpha | \Phi_i^N \rangle. \qquad (2)$$

Here the coefficients $C_i^N$ are obtained by the CCSD expansion of $|N\rangle$ in terms of Slater determinants $|\Phi_i^N\rangle$, $|N\rangle = \sum_i C_i^N |\Phi_i^N\rangle$, retaining only those coefficients with non-negligible magnitude, $|C_i^N| > 10^{-7}$, and including determinants up to double excitations. The matrix element $\langle \Phi_j^{N-1} | \hat{c}_\alpha | \Phi_i^N \rangle$ is non-zero only when the operator $\hat{c}_\alpha$ destroys an electron in the αth orbital that is occupied in $|\Phi_i^N\rangle$ but empty in $|\Phi_j^{N-1}\rangle$.

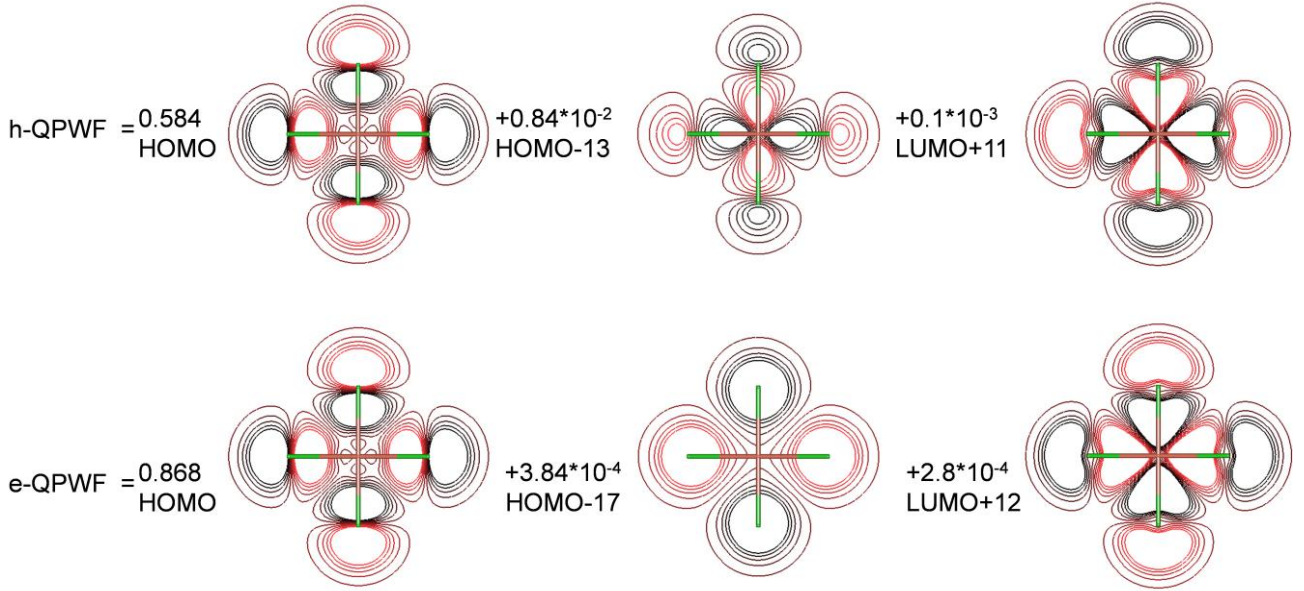

FIG. 2 (color online). Construction of the quasiparticle wave function of copper chloride. Expansion coefficients of h-QPWF (top row) and e-QPWF (bottom row) of copper chloride referred to the HF molecular orbital basis set. Only the orbital contributions with the largest weight are shown.

The construction of $\varphi(\mathbf{r})$ is illustrated in Fig. 2, showing the orbitals with the largest weight that span the h- and e-QPWF of copper chloride, as obtained by the CCSD calculation. In the absence of correlation, only the HOMO and LUMO survive with unitary weights (in Fig. 2 HOMO and LUMO coincide since the *N* state has an odd number of electrons). Correlation mixes orbitals of different energies---but like symmetries---which typically interfere destructively and reduce both weight and extension of $\varphi(\mathbf{r})$. The overall effect of such hybridization is clarified by excluding the HF contribution from the QPWF (right column of Fig. 1, h-QPWF-ZERO and e-QPWF-ZERO). It is then clear that the depletion of the h-QPWF amplitude on the Cu site is due to the superposition of SOMO-HF and h-QPWF-ZERO orbital lobes of alternate signs (black and red contour lines in Fig. 1 indicate positive and negative values, respectively).

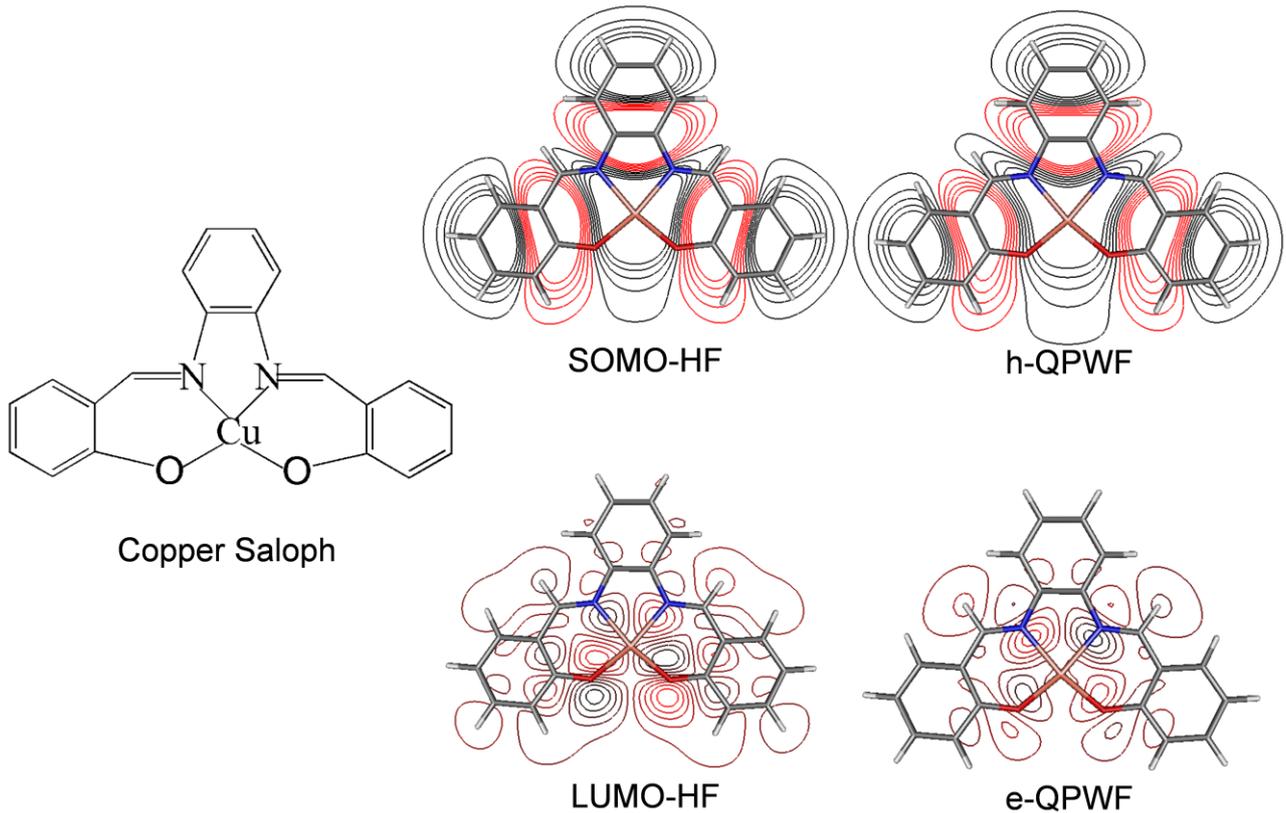

FIG. 3 (color online). Molecular orbitals and quasiparticle wave functions of copper-saloph. Contour plots of Hartree-Fock frontier orbitals (SOMO-HF and LUMO-HF, center column) and quasiparticle wave functions (h-QPWF and e-QPWF, right column) of copper-saloph. The chemical structure of the complex is schematized in the left panel. All contours are taken 4.0 Å over the molecular plane and parallel to it. Black (red) contours have positive (negative) values. The contour line maximum (minimum) value is $10^{-4}$ Å$^{-3/2}$ ( $-10^{-4}$ Å$^{-3/2}$ ). Gray, brown, red, blue, and white segments of the molecule backbone indicate the C, Cu, O, N, and H sites, respectively.

State-of-the-art STS techniques require that the investigated complexes are large enough to remain steady on the substrate during the measure. As a candidate we propose another copper-based molecule, copper-saloph [32]. The chemical structure is shown in Fig. 3, together with the HF frontier orbitals and the corresponding QPWFs (see Supplementary Discussion S5 for density functional theory predictions [33]). The SOMO-HF and LUMO-HF are different, as there are various valence configurations of similar energies that are relevant for ionization. The depicted orbitals are those that, starting from the $N$ HF configuration, are emptied or filled in the $N - 1$ and $N + 1$ states, respectively.

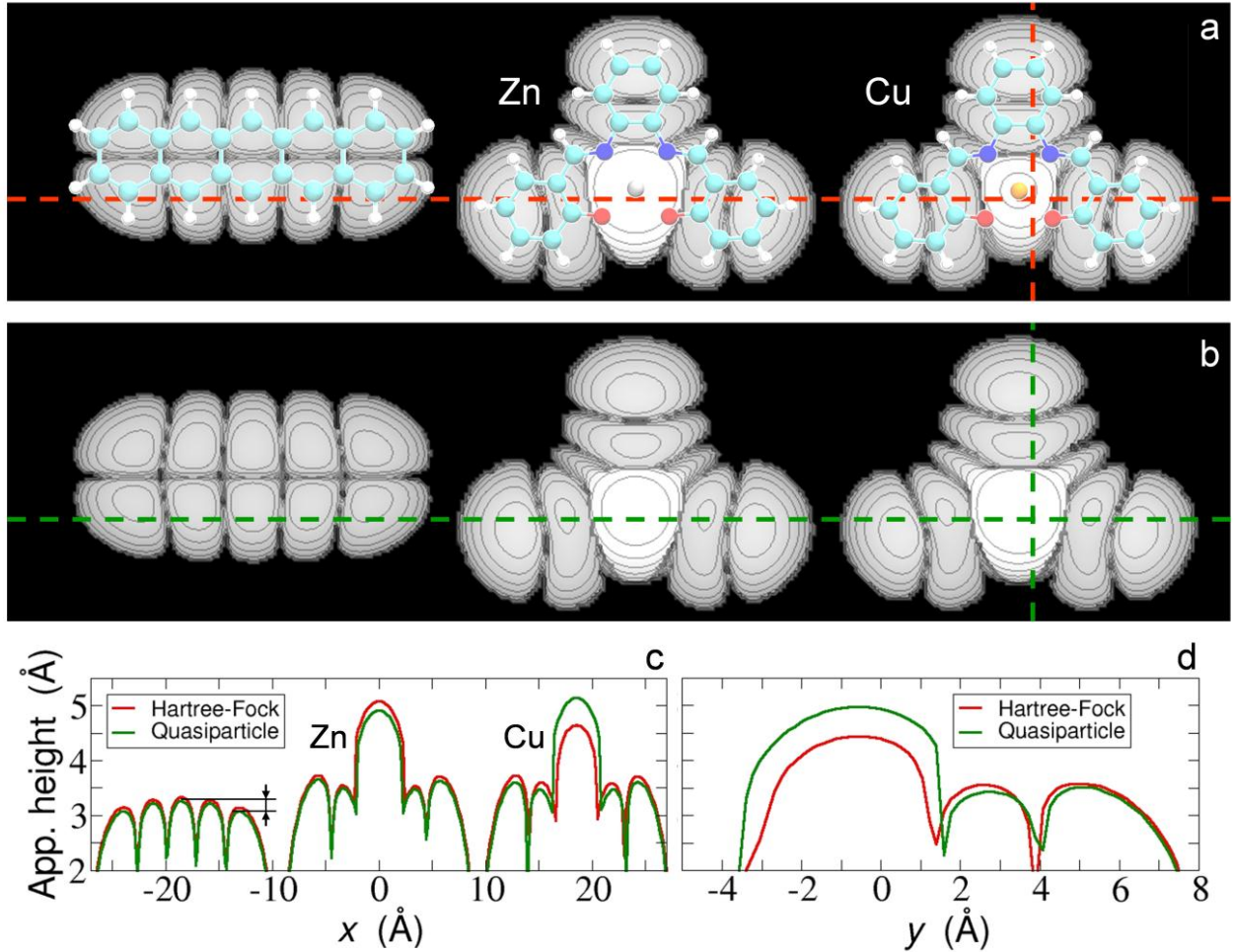

FIG. 4 (color online). Simulated constant-current STS images of positively ionized pentacene, zinc-saloph, and copper-saloph. The density threshold is $6.7 \cdot 10^{-7}$ Å$^{-3}$. (a) Contour plot in the $xy$ plane of the STS images obtained by Hartree-Fock SOMOs. The atomic structure skeletons are superimposed, with the color code: cyan=C, white=H, red=O, blue=N, gray=Zn, orange=Cu. (b) Contour plot of the STS images obtained by hole quasiparticle wave functions. The level of the theory is CISD. (c) Slice of the Hartree-Fock (red curves) and quasiparticle (green curves) STS images obtained along the horizontal dashed line shown in panels a and b. (d) Slice of the Hartree-Fock (red curve) and quasiparticle (green curve) STS images obtained along the vertical dashed line shown in panels a and b.

The quasiparticle wave functions are calculated at the level of configuration interaction with singles and doubles [30] (CISD), as the superior CCSD calculation is unattainable due to the size of the molecule. Figure 3 shows that correlation---already at the CISD level---profoundly alters the orbital images. The amplitude probability is increased on the Cu site in the h-QPWF image with respect to the SOMO-HF, extending well over the nitrogen and oxygen sites (cf. black contour lines over the blue and red segments). The LUMO-HF is qualitatively modified, due to both the strong suppression of amplitude over all external lobes and the alteration of nodal surfaces in an extended region around the oxygen sites (cf. red segments in the e-QPWF).

These changes are reflected in the predicted STS images. Both figures 4 and 5 show the STS hole maps of copper-saloph and the substituted molecule zinc-saloph, obtained in the constant-current mode, at two different current values. Zinc-saloph has the same chemical structure of copper-saloph except for Zn ion in place of Cu [cf. the atomic structure skeletons superimposed onto the images of Fig. 4(a)]. Figure 4 also includes pentacene [19] as a reference map. The values of contour lines in the $xy$ plane signify the apparent height of the molecule, i.e., the distance $z$ from the molecule plane at which the square modulus of the wave function assumes a given value (respectively $6.7 \cdot 10^{-7}$ Å$^{-3}$ in Fig. 4 and $6.7 \cdot 10^{-5}$ Å$^{-3}$ in Fig. 5). Such value is proportional to the current flowing through the STS tip if the transport energy window includes only the resonance between the ground states of the $N$ and $N + 1$ (or $N - 1$) molecule [34--36].

At large tip-molecule distance, the top of the hole quasiparticle map of copper-saloph [right image in Fig. 4(b)], located over the central Cu site, is patently higher than in the uncorrelated image [right image in Fig. 4(a)], whereas the eminence of the remaining lobes is slightly lower. On the other hand, the summit of the quasiparticle map of zinc-saloph [Fig. 4(b) center] is slightly lower than the height of the SOMO-HF [Fig. 4(a) center]. Importantly, the two predicted quasiparticle images shown in Fig. 4(b) almost coincide. Since zinc-saloph is insensitive to correlation, having only one valence configuration, one may use it as a benchmark to unveil the many-body effect.

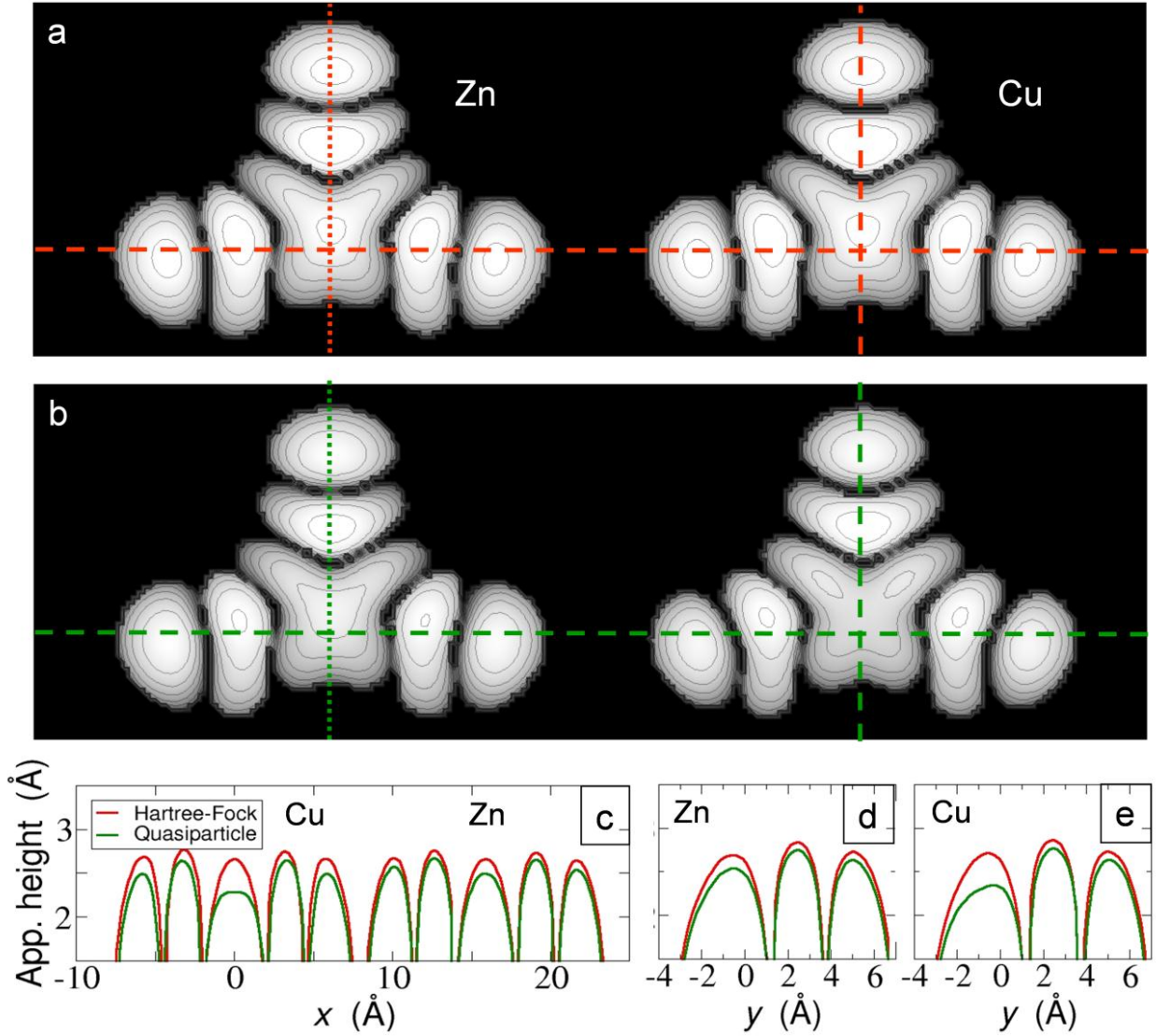

FIG. 5 (color online). Simulated constant-current STS images of positively ionized zinc-saloph and copper-saloph at closer tip-molecule distance than in Fig. 4 (the density threshold is $6.7 \cdot 10^{-5}$ Å$^{-3}$). Contour plot images of (a) Hartree-Fock SOMOs and (b) quasiparticle wave functions in the $xy$ plane. Slices of the Hartree-Fock (red curves) and quasiparticle (green curves) images obtained along the (c) horizontal and (d-e) vertical dashed lines shown in panels a and b.

This is best illustrated by cutting the contour STS maps along the $x$ axis [dashed horizontal lines in Figs. 4(a) and 4(b)]. Figure 4(c) allows to compare the hole quasiparticle profiles of copper- and zinc-saloph along $x$ (green lines) with their uncorrelated images (red lines). Correlation changes the apparent height of copper-saloph not homogeneously along the axis, with an increase of ≈ 0.5 Å over the Cu site and a slight decrease elsewhere. Therefore the contrast between the measured hole images of copper- and zinc-saloph is the fingerprint of correlation: If the top of copper-saloph is found to be comparable to the summit of zinc-saloph, one may infer that the former quasiparticle map is altered by correlation. On the contrary, in the uncorrelated case the top of copper-saloph is expected to be lower than the summit of zinc-saloph.

A residual ambiguity is that copper- and zinc-saloph images could be identical as an artifact of poor STS resolution. A way to exclude this is to further compare the hole quasiparticle images with that of the thoroughly investigated [11,12,14] pentacene, that is unaffected by correlation [left images in Figs. 4(a-c)]. Consider the pentacene image profile in Fig. 4(c): If the measurement is able to resolve the height difference of ≈ 0.2 Å between the central and outer pentacene peaks (highlighted by the arrows), then the sensitivity is larger than that required (≈ 0.5 Å) to assess the identity of copper- and zinc-saloph images. Lateral lobes of both molecules, being minimally affected by correlation, may be used as intramolecular height references. In summary, the comparison among the measured hole images of pentacene, copper- and zinc-saloph at large tip-molecule distance provides a clear-cut experimental signature of the many-body effect.

Another striking feature of correlation emerges as the tip-molecule distance is reduced (Fig. 5). The trend is reversed with respect to Fig. 4, as the uncorrelated images of copper- and zinc-saloph are now identical [cf. red curves in the STS map cuts along $x$ and $y$ in Figs. 5(c-e)], but the quasiparticle peak over the Cu site is lower than that of Zn (green curves). Again, zinc-saloph is only slightly affected by correlation, as its HF and quasiparticle maps almost overlap, whereas copper-saloph experiences a drastic reduction of weight, the height difference between HF and quasiparticle profiles over the Cu ion being around 0.5 Å [Figs. 5(c) and 5(e)].

The overall behavior of copper-saloph images over the range of tip-molecule distances explored in Figs. 4 and 5 has a clear physical meaning: Coulomb interaction beyond mean field transfers the quasiparticle weight from the neighborhood of Cu ion into the outer vacuum region, creating a correlation hole. This novel effect may be checked most easily by scanning images at different current values: The correlated images of copper- and zinc-saloph are expected to overlap at low current, whereas Cu is lower than Zn in high-current maps. The opposite holds for the uncorrelated images, which are identical at high tip-molecule currents but part at low currents. Further evidence is provided by the negative-ionization images (Supplementary Discussion 1 [33]).

In the Supplementary Discussion 3 [33] we study the interaction between copper-saloph and a typical insulating substrate, a NaCl monolayer [11--13,15]. The coupling turns out to be mainly of electrostatic nature, except for the Cu atom, that binds to the $Cl^-$ ion of the monolayer, distorting the STS image of the isolated molecule. Nevertheless, the differences between quasiparticle and HF images remain significant and measurable. Moreover, in the Supplemental Material [33] we show that our STS results are insensitive to the screening effect of the backgate (Discussion S4) and robust against modifications in the molecule that do not involve the metal center (Discussion S6). Overall, we find significant many-body effects, that remain large and measurable even when the extrinsic effect of the substrate is taken into account.

In conclusion, by obtaining the quasiparticle wave function through quantum-chemical methods, we have demonstrated that many-body effects may profoundly affect the STS images of molecules with metal centers. Our results highlight the still unexplored potential of STS as a probe of electron correlation in molecules and nano-objects. So far, the importance of electron correlation in molecules has been inferred indirectly through structural and energetic properties (e.g., bond energies, geometries, optical spectra). Our work shows that STS directly images the effects of electron correlation on the wave function. We hope that our findings may stimulate further STS experiments along this path.

The authors thank Germar Hoffmann, Peter Liljeroth, and Elisa Molinari for useful discussions. Funding from INFM under the Young Researcher Seed Project 2008 initiative and Fondazione Cassa di Risparmio di Modena under the project COLDandFEW is gratefully acknowledged. Computer time has been provided by CINECA under the CINECA-ISCRA supercomputer project grants IscrB_FERMIFEW, IscrC_FEW1D, IscrC_QUASIPAR.
*Corresponding author. Electronic address: massimo.rontani@nano.cnr.it

†Electronic address: stefano.corni@nano.cnr.it

[1] R. Wiesendager, *Scanning Probe Microscopy and Spectroscopy* (Cambridge University Press, Cambridge, 1994).

[2] W. A. Hofer, A. S. Foster, and A. L. Shluger, Rev. Mod. Phys. **75,** 1287-1331 (2003).

[3] M. Tsukada, K. Kobayashi, N. Isshiki, and H. Kageshima, Surf. Science Rep. **13,** 267-304 (1991).

[4] L. P. Kouwenhoven, C. M. Marcus, P. L. McEuen, S. Tarucha, R. M. Westervelt, and N. S. Wingreen, in *Mesoscopic Electron Transport*, edited by L. L. Sohn, L. P. Kouwenhoven, and G. Schoen (Kluwer, Dordrecht, 1997), p. 105-214.

[5] M. A. Reed, C. Zhou, C. J. Muller, T. P. Burgin, and J. M. Tour, Science **278,** 252-254 (1997).

[6] L. C. Venema, J. W. G. Wildöer, J. W. Janssen, S. J. Tans, H. L. J. Temminck Tuinstra, L. P. Kouwenhoven, and C. Dekker, Science **283,** 52-55 (1999).

[7] B. Grandidier, Y. M. Niquet, B. Legrand, J. P. Nys, C. Priester, D. Stiévenard, J. M. Gérard, and V. Thierry-Mieg, Phys. Rev. Lett. **85,** 1068 (2000).

[8] E. E. Vdovin, A. Levin, A. Patanè, L. Eaves, P. C. Main, Y. N. Khanin, Y. V. Dubrovskii, M. Henini, and G. Hill, Science **290,** 122-124 (2000).

[9] O. Millo, D. Katz, Y. W. Cao, and U. Banin, Phys. Rev. Lett. **86,** 5751 (2001).

[10] T. Maltezopoulos, A. Bolz, C. Meyer, C. Heyn, W. Hansen, M. Morgenstern, and R. Wiesendanger, Phys. Rev. Lett. **91,** 196804 (2003).

[11] J. Repp, G. Meyer, S. M. Stojkovic, A. Gourdon, and C. Joachim, Phys. Rev. Lett. **94,** 026803 (2005).

[12] A. Bellec, F. Ample, D. Riedel, G. Dujiardin, and C. Joachim, Nano Lett. **9,** 144-147 (2009).

[13] C. J. Villagomez, T. Zambelli, S. Gauthier, A. Gourdon, S. Stojkovic, and C. Joachim, Surf. Sci. **603,** 1526-1532 (2009).

[14] W.-H. Soe, C. Manzano, A. De Sarkar, N. Chandrasekhar, and C. Joachim, Phys. Rev. Lett. **102,** 176102 (2009).

[15] P. Liljeroth, I. Swart, S. Paavilainen, J. Repp, and G. Meyer, Nano Lett. **10,** 2475-2479 (2010).

[16] M. Rontani and E. Molinari, Phys. Rev. B **71,** 233106 (2005).

[17] G. Maruccio, M. Janson, A. Schramm, C. Meyer, T. Matsui, C. Heyn, W. Hansen, R. Wiesendanger, M. Rontani, and E. Molinari, Nano Lett. **7,** 2701 (2007).

[18] G. Bester, D. Reuter, L. He, A. Zunger, P. Kailuweit, A. D. Wieck, U. Zeitler, J. C. Maan, O. Wibbelhoff, and A. Lorke, Phys. Rev. B **76,** 075338 (2007).

[19] D. Toroz, M. Rontani, and S. Corni, J. Chem. Phys. **134,** 024104 (2011).

[20] A. Secchi and M. Rontani, Phys. Rev. B **85,** 121410(R) (2012).

[21] D. Goldhaber-Gordon, H. Shtrikman, D. Mahalu, D. Abusch-Magder, U. Meirav, and M. A. Kastner, Nature **391,** 156-159 (1998).

[22] S. M. Cronenwett, T. H. Oosterkamp, and L. P. Kouwenhoven, Science **281,** 540-544 (1998).

[23] J. Park, A. N. Pasupathy, J. I. Goldsmith, C. Chang, Y. Yaish, J. R. Petta, M. Rinkoski, J. P. Sethna, H. D. Abruña, P. L. McEuen, and D. C. Ralph, Nature **417,** 722-725 (2002).

[24] W. Liang, M. P. Shores, M. Bockrath, J. R. Long, and H. Park, Nature **417,** 725-729 (2002).

[25] N. Maire, F. Hohls, T. Ludtke, K. Pierz, and R. J. Haug, Phys. Rev. B **75,** 233304 (2007).

[26] J. B. Neaton, M. S. Hybertsen, and S. G. Louie, Phys. Rev. Lett. **97,** 216405 (2006).

[27] K. S. Thygesen and A. Rubio, Phys. Rev. Lett. **102,** 046802 (2009).

[28] C. Freysoldt, P. Rinke, and M. Scheffler, Phys. Rev. Lett. **103,** 056803 (2009).

[29] P. Fulde, *Electron Correlations in Molecules and Solids* (Springer, Berlin, 1995).

[30] T. Helgaker, P. Jørgensen, and J. Olsen, *Molecular Electronic Structure Theory* (Wiley, Chichester, 2000).

[31] M.J. Frisch, G. W. Trucks, H.B. Schlegel, et al. Gaussian 09, Revision B.01 Gaussian, Inc., Wallingford CT, 2010.

[32] E. Suresh, M. M. Bhadbhade, and D. Srinivas, Polyhedron **15,** 4133-4144 (1996).

[33] See Supplemental Material at XXX for STS electron maps of copper- and zinc-saloph (S1), the dependence of wave function tail on the basis set (S2), the effects of the insulating substrate (S3) and backgate image charges (S4) on STS maps, copper-saloph STS images from density functional theory (S5), quasiparticle wave functions and STS images of copper-(deh-salen) (S6).

[34] We have obtained the constant-current STS maps, $s(x,y)$, by arbitrarily choosing a set-point threshold $t$ and then defining $s(x,y)=z$ with $z$ such that $|\varphi(\mathbf{r})|^2 = t$. As the value of the STS current is proportional to $|\varphi(\mathbf{r})|^2$ [19], choosing $t$ is equivalent to choose a set point current. We have performed the calculations of constant-current STS images with an in-house code.

[35] Here we adopt the common Tersoff-Hamman model of the STS tip as a *s*-wave perturbation [1,19]. For atomic-like resolution it may be required to consider the microscopic structure of the tip, so STS images are convolutions of tip and molecule states [2,11,37].

[36] Since states of similar energy might not be resolved individually by STS [20], we have inspected the excitation spectrum of copper-saloph in the final charge state (both $N - 1$ and $N + 1$). As this calculation is prohibitive at the CI level, we have reverted to the time dependent HF scheme. We have found that the lowest-lying excited state (for $N - 1$) is 0.5 eV higher in energy than the ground state, allowing for full STS resolution.

[37] S. Sobczyk, A. Donarini, and M. Grifoni, Phys. Rev. B **85,** 205408 (2012).

Supplemental Material

# Proposed alteration of images of molecular orbitals obtained using a scanning tunnelling microscope as a probe of electron correlation

Dimitrios Toroz, Massimo Rontani*, Stefano Corni

*CNR-NANO Research Center S3, Via Campi 213a, 41125 Modena, Italy*

S1. SIMULATED STS IMAGES OF NEGATIVELY-IONIZED ZINC-SALOPH AND COPPER-SALOPH

S2. ANALYZING POSSIBLE ARTIFACTS IN STS IMAGES DUE TO THE USAGE OF A GAUSSIAN BASIS SET

S3. EFFECTS OF THE INSULATING SUBSTRATE ON STS IMAGES

S4. EFFECTS OF BACKGATE-INDUCED IMAGE CHARGES ON STS MAPS

S5. DENSITY FUNCTIONAL THEORY DESCRIPTION OF COPPER-SALOPH STS IMAGES

S6. QUASIPARTICLE WAVE FUNCTIONS AND STS IMAGES OF COPPER-(DEH-SALEN)

# S1. SIMULATED STS IMAGES OF NEGATIVELY-IONIZED ZINC-SALOPH AND COPPER-SALOPH

In this section we present the predicted STS images of negatively-ionized copper-saloph and zinc-saloph, obtained in the constant-current mode. The methodology used to obtain the STS maps is the same as that employed for Figs. 4 and 5 of main text. Correlation strongly affects the electron-like excitation of copper-saloph, as shown in Fig. S1. In view of a possible experimental validation, the key findings are: (i) The intricate topology of nodal surfaces changes qualitatively from the HF [Fig. S1(a)] to the quasiparticle [Fig. S1(b)] image, whereas zinc-saloph is insensitive to many-body effects. (ii) The cuts of copper-saloph contour map along the axes [Figs. S1(c) and S1(d)] reveal a huge increase of the quasiparticle weight in the central region, so the map top is shifted upwards of ≈ 2 Å and becomes comparable to the summit of zinc-saloph. (iii) The quasiparticle profile exhibits a significant lateral squeeze of ≈ 1 Å along $x$ [Fig. S1(c)].

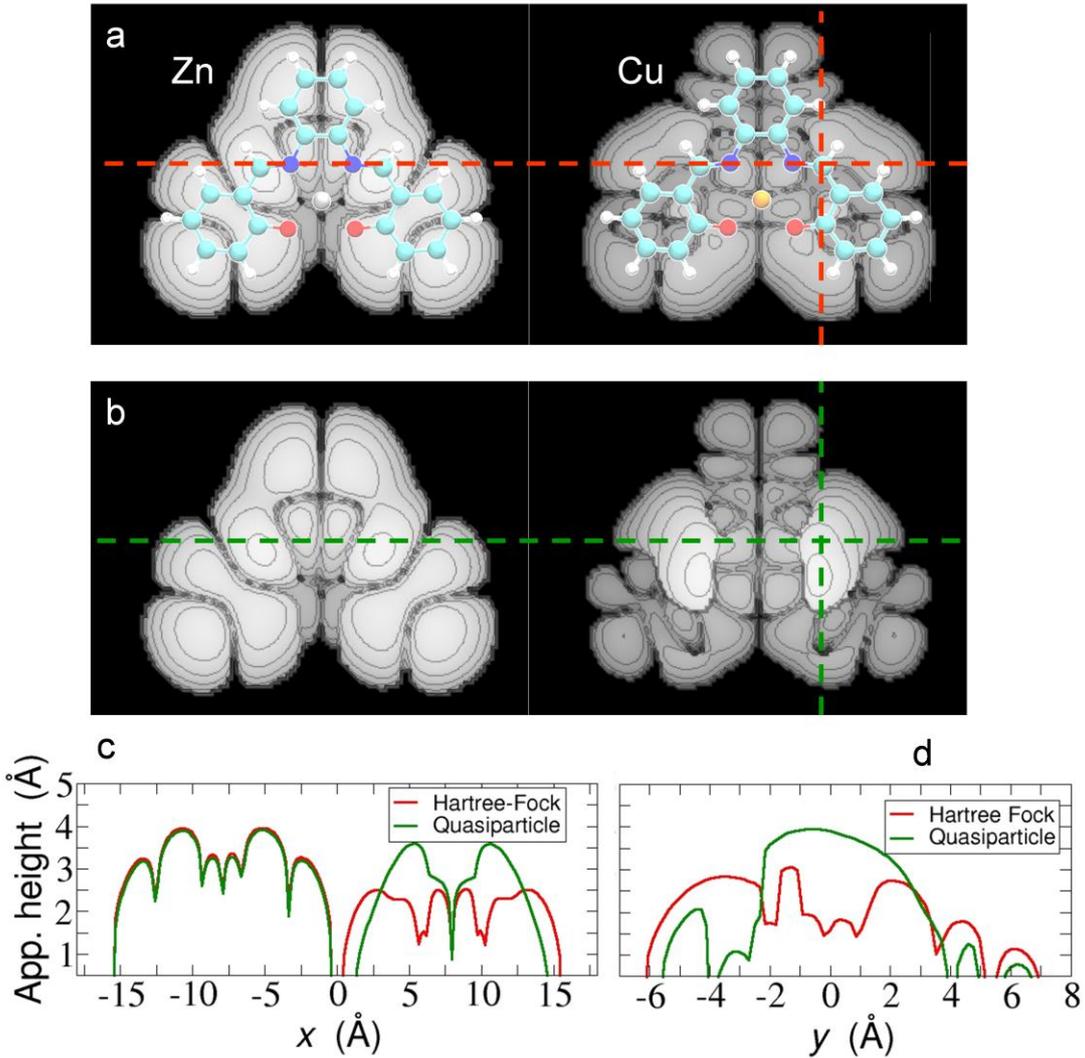

FIG. S1. Simulated constant-current STS images of negatively ionized zinc-saloph and copper-saloph. (a) Contour plot in the $xy$ plane of the STS images obtained by Hartree-Fock LUMOs. (b) Contour plot of the STS images obtained by electron quasiparticle wave functions. (c) Slice of the Hartree-Fock (red curves) and quasiparticle (green curves) STS images obtained along the horizontal dashed line shown in panels a and b. (d) Slice of the Hartree-Fock (red curve) and quasiparticle (green curve) STS images obtained along the vertical dashed line shown in panels a and b.

## S2. ANALYZING POSSIBLE ARTIFACTS IN STS IMAGES DUE TO THE USAGE OF A GAUSSIAN BASIS SET

A single Gaussian function has an incorrect asymptotic behavior and thus may give rise to artifacts in STS images evaluated at large distances from the molecular plane. Nevertheless, linear combinations of nodeless Gaussian functions are a basis set for continuous orbital functions [T. Helgaker, P. Jørgensen, and J. Olsen, *Molecular Electronic Structure Theory* (Wiley, Chichester, 2000)] and thus can, in principle, provide a realistic exponential decay for the wave function. Being a variational approach, increasing the size of the Gaussian basis set improves the quality of the wave function tail as compared to the exact exponential decay.

To check the behavior of the wave function decay obtained from the Gaussian basis set used in this work (Lanl2DZ), we have performed HF calculations on copper-saloph and zinc-saloph with a larger basis set (Ahlrichs' TZV + an extra set of $p$ diffuse functions). Our aim here is to investigate how the tail of the wave function square modulus evaluated at large distances from the molecular plane depends on the size of the basis set.

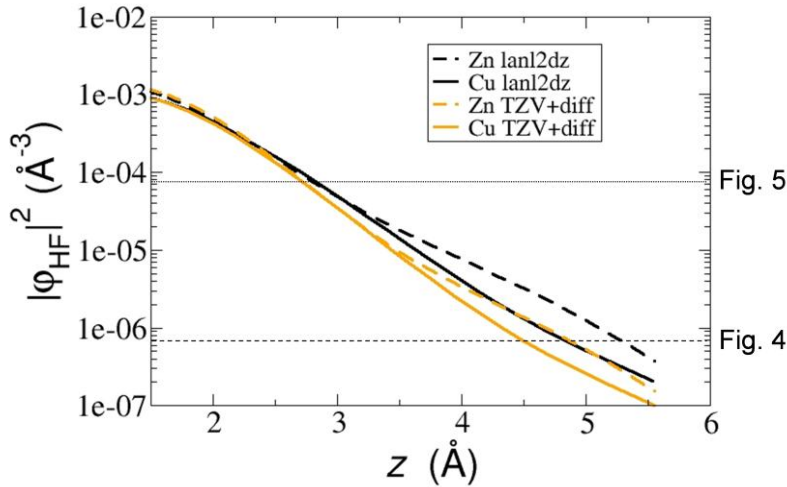

FIG. S2. Decay of the HF SOMO square modulus for copper-saloph and zinc-saloph with different basis sets. The decay is plotted along the direction perpendicular to the molecule, on top of either Cu or Zn atom. The label "lanl2dz" refers to the basis set used for all the calculations in this work, whereas "TZV+diff" refers to the calculation with Ahlrichs' TZV basis set supplemented by a diffuse $p$ shell on Cu and Zn. The horizontal lines mark the thresholds used for the plots of Figs. 4 and 5 in the main text (black curves).

The results for the HF SOMO decay in the direction perpendicular to the copper-saloph (zinc-saloph) plane, taken on top of Cu (Zn), are reported in Fig. S2. It is clear that up about 3Å from the molecule the two basis sets give the same results, for both Cu and Zn. It is also noticeable that in the region around 3Å the wave function is decaying exponentially, i.e., linearly in the log scale of the plot. The contour plots in Fig. 3 and the STS images in Fig. 5 in the main text are within this regime (the density threshold around $10^{-4}$ Å$^{-3}$ used to create Fig. 5 is highlighted by a dotted line in Fig. S2).

For distances larger than 3 Å, an increasing discrepancy between the two basis sets is seen. Nevertheless, both sets predict Zn to decay slower than Cu, which is responsible for the qualitative features seen in Fig. 4 (the density threshold used to create Fig. 4 is highlighted by a dashed line in Fig. S2). Therefore we infer that, whereas the absolute values may be rather approximated in the distance range of Fig. 4, the relative comparison between Zn and Cu is still meaningful.

In conclusion, on top of the Cu/Zn atom (the region more important for comparison) the Gaussian basis set that we have employed is quantitatively reproducing the wave function shape up to 3Å, and it is able to qualitatively reproduce differences between copper-saloph and zinc-saloph at larger distances as well.

Whereas it is not possible to perform CI calculations with the larger basis set, the qualitative effects of correlations discussed in the main text should also be robust. In fact, correlation is actually exerting its effects via the regions of high electron density, where the wave function is well described by the smaller basis set.

## S3. EFFECTS OF THE INSULATING SUBSTRATE ON STS IMAGES

To verify whether the presence of a substrate may hide correlation effects on STS images of copper-saloph, we have considered this molecule on a typical experimental substrate, NaCl(100). We have relaxed the structure of the molecule at the HF level, using a QM/MM approach (see details below) obtaining the geometry that is plotted in Fig. S3.

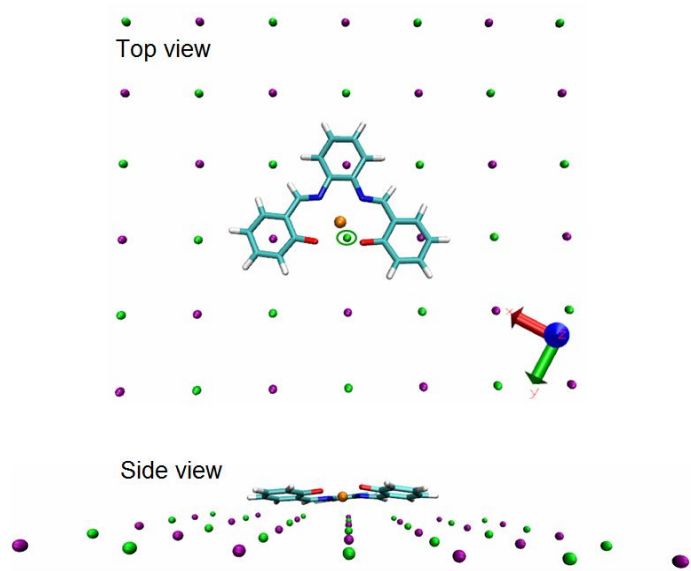

FIG. S3. Geometric arrangement of copper-saloph on a NaCl(100) monolayer. The geometry has been optimized at the Hartree-Fock level, using a QM description of copper-saloph and the closest $Cl^-$ ion (circled in the Top view panel) and a MM description of the rest of the NaCl monolayer. Color code: green=Cl, purple=Na, cyan=C, white=H, red=O, blue=N, orange=Cu.

The distortion from planarity of the molecular geometry is due to the interaction of Cu with the underlying Cl$^-$ ion. NaCl, therefore, is not an ideal substrate for this class of molecules, as the non-planar arrangement and the coupling with the substrate make the interpretation of STS results more complex. Nevertheless, NaCl provides a convenient test to check whether correlation effects in STS survive to relatively strong interactions with the substrate.

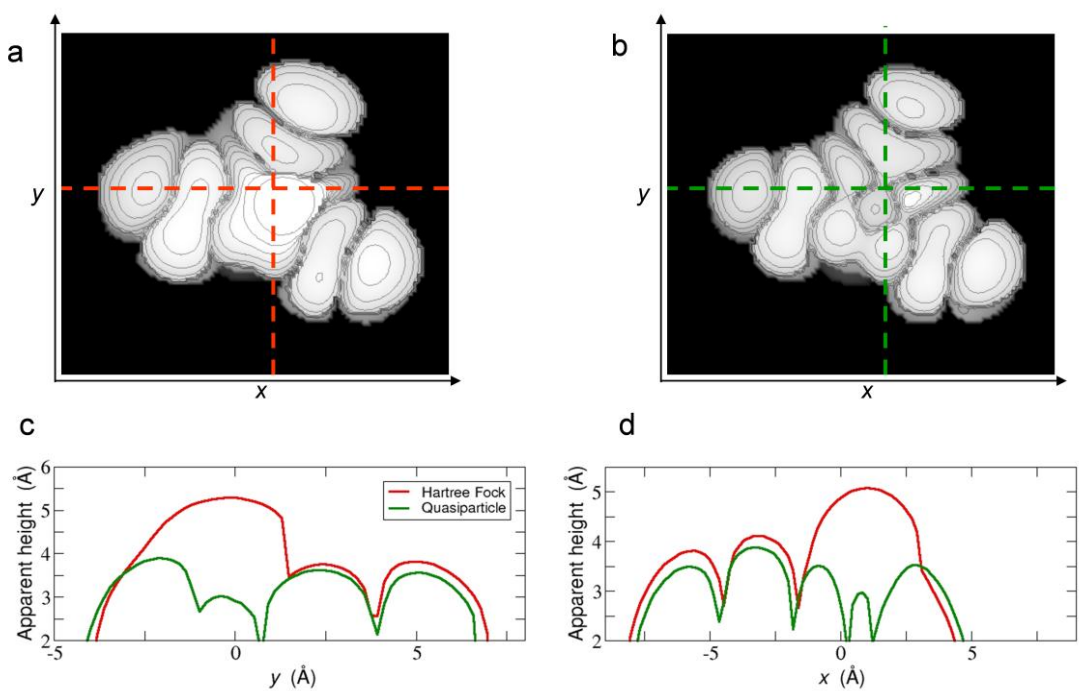

FIG. S4. Simulated constant-current STS images of positively ionized copper-saloph on an NaCl(100) monolayer. (a) Contour plot in the *xy* plane of the STS image obtained by the Hartree-Fock SOMO. (b) Contour plot of the STS image obtained by the hole quasiparticle wave function. The level of the theory is CISD for copper-saloph and the closest Cl$^-$ ion, MM for the rest. (c) Slice of the Hartree-Fock (red curve) and quasiparticle (green curve) STS images obtained along the vertical dashed line shown in panels a and b. (d) Slice of the Hartree-Fock (red curve) and quasiparticle (green curve) STS images obtained along the horizontal dashed line shown in panels a and b.

The inspection of Fig. S4 shows that this is indeed the case: the STS maps of copper-saloph on the substrate calculated respectively from Hartree Fock SOMO [Fig. S4(a)] and from the hole quasiparticle wave function [Fig. S4(b)] show clear differences. In particular, the central region, around the Cu center, is much higher for the uncorrelated STS than for the correlated one. Height differences amount to approximately 2 Å. Here, the uncorrelated STS map is the highest, whereas for the isolated molecule it was the opposite at large distance from the molecule (Fig. 4) and the same at smaller distance (Fig. 5).

*Details on the calculation*: copper-saloph has been placed on a single NaCl(100) layer composed of 12 Na$^+$ and 12 Cl$^-$ ions. A QM/MM procedure has been used, where copper-saloph plus the Cl$^-$ ion closest to the Cu has been included in the QM part, whereas the rest of the NaCl layer is treated MM as a set of point charges (+1 for Na$^+$ and -1 for Cl$^-$). Geometry optimization has been performed at the HF level, freezing the NaCl layer. Different optimizations have been started by different initial positions, converging to qualitatively similar structures (the most stable was chosen for the calculations). To calculate quasiparticle wave functions (obtained at the CISD level), the same procedure described in the main text has been used.

## S4. EFFECTS OF BACKGATE-INDUCED IMAGE CHARGES ON STS MAPS

In the STS images of the molecule on a NaCl substrate in Figs. S4 and S11, both the structural and the electrostatics effects produced by the $Na^+$ and $Cl^-$ ions have been taken into account. The goal of this section is to verify the possible effects on the wave function of the image interaction with the metal substrate underlying the NaCl layer, which acts as a backgate in the experimental setups. To this aim, we have performed calculations on copper-saloph by including interaction with image charges explicitly in the Hartree-Fock Hamiltonian by the method described in S. Corni and J. Tomasi, J. Chem. Phys. **114,** 3739 (2001); S. Corni, J. Phys. Chem. B **109,** 3423 (2005). This method treats the metal as a perfect conductor via classical electrostatics, which was shown to give results of quality comparable to GW for the quasiparticle band gap in J. B. Neaton, M. S. Hybertsen, and S. G. Louie, Phys. Rev. Lett. **97,** 216405 (2006). Computational details are given below.

The profiles along $x$ and $y$ of the STS constant-current images obtained in the presence of the metal are reported in Fig. S5. They are clearly indistinguishable from those in vacuo, showing that image interaction with the underlying metal substrate does not affect STS maps.

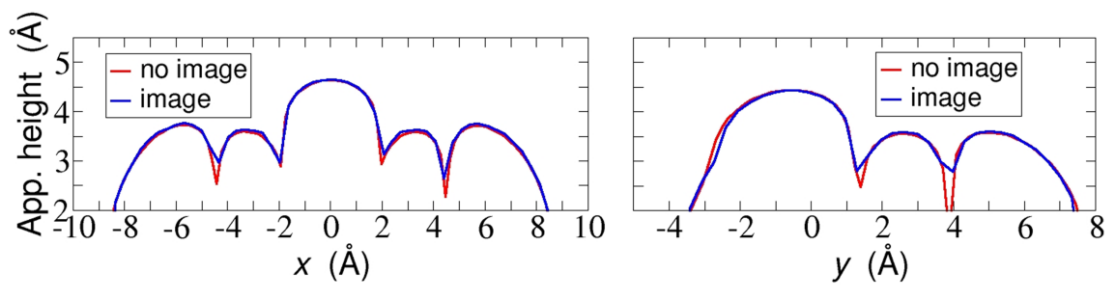

FIG. S5. Slices of constant-current STS maps of positively ionized copper-saloph obtained with (blue curves) and without (red curves) accounting for the image interaction with the metal substrate underlying the NaCl layer. The slices are taken as in Fig. 4 of the main text. Results have been obtained at the Hartree Fock level.

*Computational details.* Geometry and basis set were the same used for in vacuo calculations. The metal image plane was chosen to be 11.5 Å from the surface plane, that roughly corresponds to the molecule-metal distance imposed by 3 NaCl layers, as usable in the experiments [J. Repp, G. Meyer, S. M. Stojkovic, A. Gourdon, and C. Joachim, Phys. Rev. Lett. **94,** 026803 (2005)]. The numerical method applied to include the image interaction exploits an Integral Equation Formalism - Boundary Element Method resolution of the electrostatic problems, as detailed in S. Corni, J. Phys. Chem. B **109,** 3423 (2005). Specifically, a boundary surrounding the molecule has been built as a union of spheres centered on each molecule atom (but hydrogens) with radius 3.2 Å. Calculations have been performed with a local version of GAMESS.

# S5. DENSITY FUNCTIONAL THEORY DESCRIPTION OF THE COPPER-SALOPH STS IMAGES

Density functional theory (DFT) is often able to recover correlation effects on molecular quantities to a degree similar to more expensive post-HF methods [see e.g. F. Jensen, *Introduction to Computational Chemistry* (Wiley, Chichester, 1999)]. To verify whether DFT is also able to reproduce correlation effects in STS images of copper-saloph, we have calculated the STS image from the DFT LUMO and compared with HF and quasiparticle CISD results. This is done in Fig. S6 for the widely used B3LYP exchange-correlation functional.

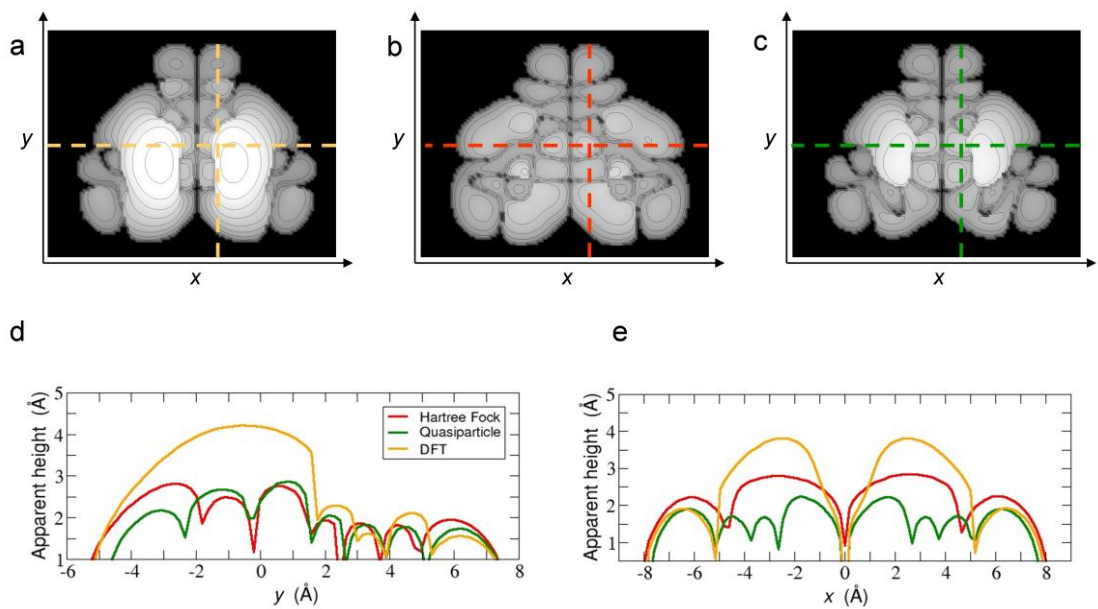

FIG. S6. Simulated constant-current STS images of negatively ionized copper-saloph including DFT results. (a) Contour plot in the *xy* plane of the STS image obtained by the DFT (B3LYP) LUMO. (b) Contour plot in the *xy* plane of the STS image obtained by the Hartree-Fock LUMO. (c) Contour plot of the STS image obtained by the electron quasiparticle wave function. The level of the theory is CISD. (d) Slice of the DFT (orange curve), Hartree-Fock (red curve), and quasiparticle (green curve) STS images obtained along the vertical dashed line shown in panels a, b, and c. (e) Slice of the DFT (orange curve), Hartree-Fock (red curve), and quasi-particle (green curve) STS images obtained along the horizontal dashed line shown in panels a, b, and c.

It is clear that the DFT is qualitatively catching the correlation effects, but it is overestimating them as it yields broader and higher lobes than those from CISD. On the contrary, analyzing DFT results for SOMO (results not shown) we found a minimal difference with HF. DFT seems thus to give erratic results. To account for correlation effects in STS images of copper-saloph, post-HF methods are therefore required.

*Details of the calculations.* The geometry of copper-saloph was the same used for HF and CISD calculations, obtained as described in the main text. The same basis set was used, with B3LYP as the exchange correlation functional.

## S6. QUASIPARTICLE WAVE FUNCTIONS AND STS IMAGES OF COPPER-(DEH-SALEN)

To verify whether the STS results are robust with respect to modifications in the molecule that do not involve the metal center, we have considered a Cu complex with the dehydrogenated salen ligand [copper-(deh-salen), see Fig. S7].

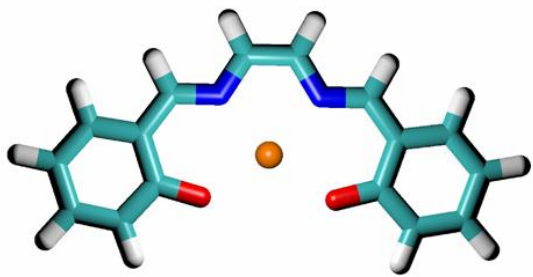

FIG. S7. The structure of copper-(deh-salen). The geometry has been optimized at the Hartree-Fock level. Color code: cyan=C, white=H, red=O, blue=N, orange=Cu.

For this molecule, slightly smaller than copper-saloph, it was possible to perform CCSD calculations. The STS images from the HF SOMO and the corresponding hole quasiparticle wave function are reported in Fig. S8, together with those of a reference molecule that we studied previously [D. Toroz, M. Rontani, and S. Corni, J. Chem. Phys. **134,** 024104 (2011)], divinyl benzene. The images obtained from the HF LUMO and the electron quasiparticle wave function are reported in Fig. S9.

At odds with copper-saloph, here electron and hole STS images are qualitatively similar, pointing at a somewhat smaller effect of correlation. Nevertheless, the differences between correlated and uncorrelated STS images, revealed by panels c and d of Figs. S8 and S9, are similar to those evidenced for copper-saloph in Fig. 4. Height differences between correlated and uncorrelated STS images are around 0.5 Å.

On the other hand, the HF and quasiparticle maps of divinyl benzene are identical (Figs. S8 and S9). Therefore, one may use divinyl benzene as a reference, by comparing its height, unperturbed by correlation, to the height of copper-(deh-salen), sensitive to many-body effects. The choice of the small divinyl benzene complex here has illustrative purposes, since there is a broad class of molecules that are unaffected by correlations [D. Toroz, M. Rontani, and S. Corni, J. Chem. Phys. **134,** 024104 (2011)], including pentacene that has been thoroughly investigated.

We have also investigated the effects of a NaCl(100) substrate on copper-(deh-salen), in the same spirit of the copper-saloph test. The QM/MM optimized geometry of copper-(deh-salen) on a monolayer of NaCl(100) is reported in Fig. S10.

As found for copper-saloph, there is a non-negligible interaction between Cu and $Cl^-$, that results in a deviation of the molecular geometry from planarity. The resulting STS images for the positively ionized molecule are reported in Fig. S11.

The difference between uncorrelated and correlated results is striking, and qualitatively similar to those found for copper-saloph on NaCl(100). Height differences of 1-2 Å between HF and CCSD results are obtained. These findings contribute supporting that correlation effects in STS images can be revealed despite an interacting substrate.

*Details of the calculations.* The protocols were the same used for copper-saloph, the only difference being that CCSD was used for the isolated copper-(deh-salen) instead of CISD.

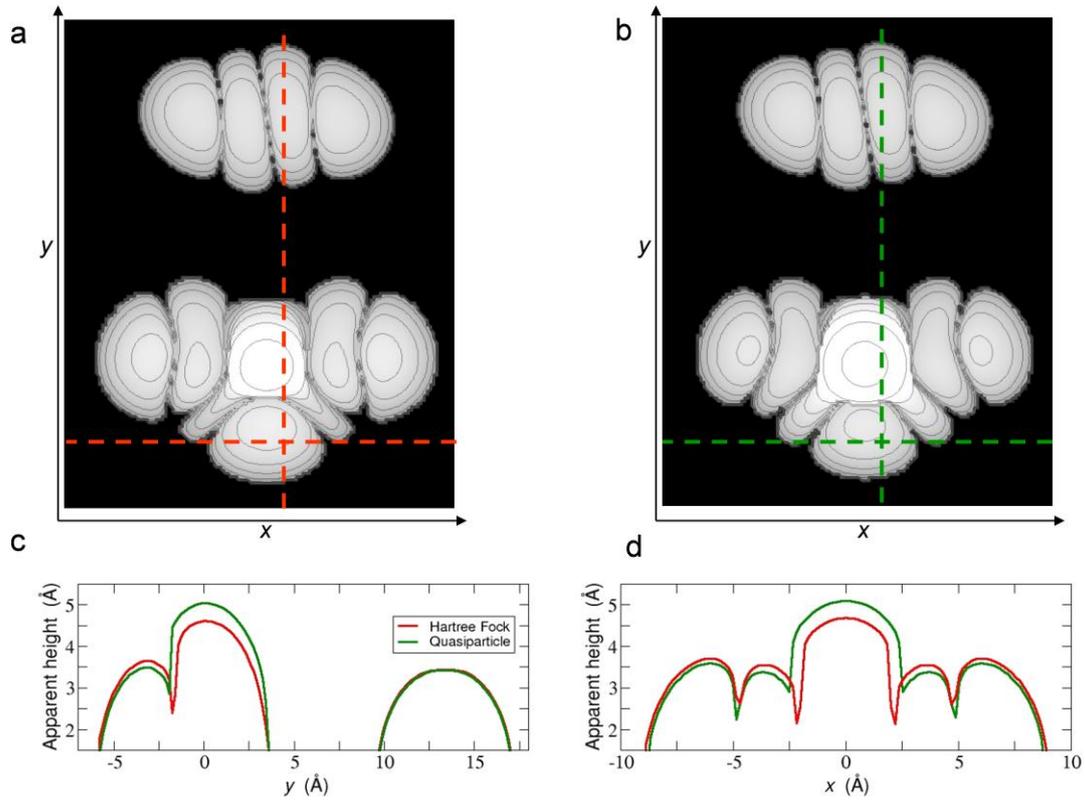

FIG. S8. Simulated constant-current STS images of positively ionized copper-(deh-salen) and divinyl benzene. (a) Contour plot in the *xy* plane of the STS images obtained by Hartree-Fock SOMOs. (b) Contour plot of the STS images obtained by hole quasiparticle wave functions. The level of the theory is CCSD for both copper-(deh-salen) and divinyl benzene. (c) Slice of the Hartree-Fock (red curve) and quasiparticle (green curve) STS images obtained along the vertical dashed line shown in panels a and b. (d) Slice of the Hartree-Fock (red curve) and quasiparticle (green curve) STS images obtained along the horizontal dashed line shown in panels a and b.

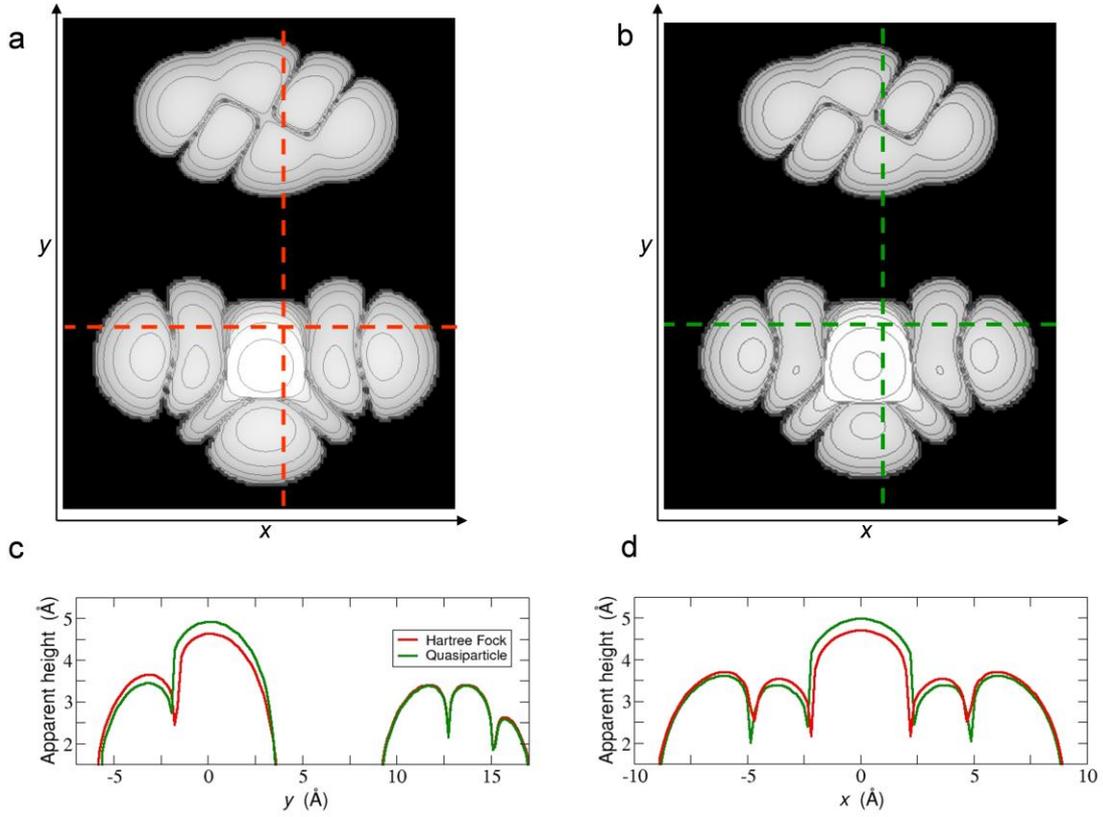

FIG. S9. Simulated constant-current STS images of negatively ionized copper-(deh-salen) and divinyl benzene. (a) Contour plot in the *xy* plane of the STS images obtained by Hartree-Fock LUMOs. (b) Contour plot of the STS images obtained by electron quasiparticle wave functions. The level of the theory is CCSD for both copper-(deh-salen) and divinyl benzene. (c) Slice of the Hartree-Fock (red curve) and quasiparticle (green curve) STS images obtained along the vertical dashed line shown in panels a and b. (d) Slice of the Hartree-Fock (red curve) and quasiparticle (green curve) STS images obtained along the horizontal dashed line shown in panels a and b.

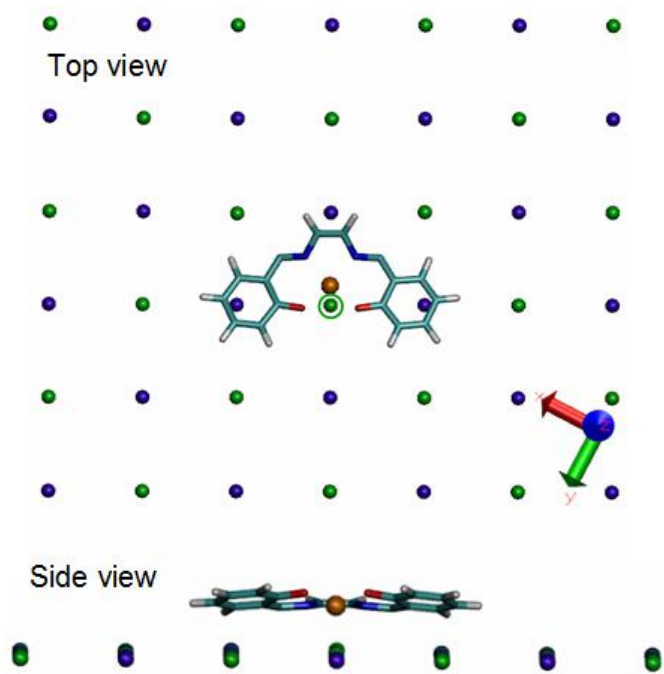

FIG. S10. Geometric arrangement of copper-(deh-salen) on a NaCl(100) monolayer. The geometry has been obtained by optimization at the Hartree-Fock level, using a QM description of copper-(deh-salen) and the closest Cl$^-$ ion (circled in the upper panel) and a MM description of the rest of the NaCl monolayer. Color code: green=Cl, violet=Na, cyan=C, white=H, red=O, blue=N, orange=Cu.

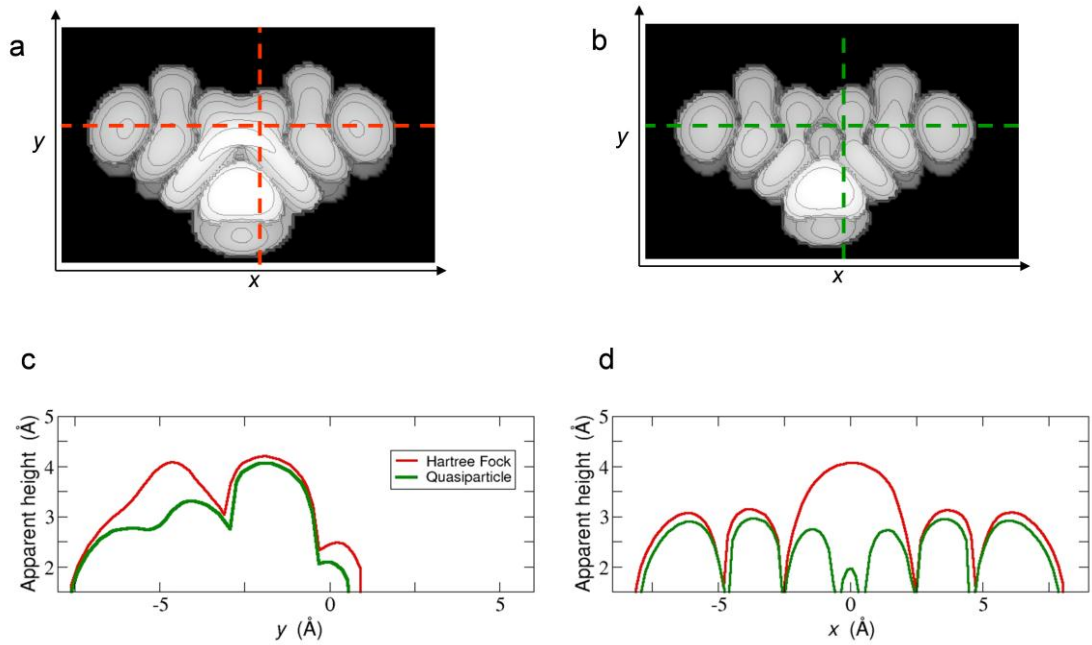

FIG. S11. Simulated constant-current STS images of positively ionized copper-(deh-salen) on an NaCl(100) monolayer. (a) Contour plot in the *xy* plane of the STS images obtained by Hartree-Fock SOMOs. (b) Contour plot of the STS images obtained by hole quasiparticle wave functions. The level of the theory is CISD for copper-saloph and the closest $Cl^-$ ion, MM for the rest. (c) Slice of the Hartree-Fock (red curve) and quasiparticle (green curve) STS images obtained along the vertical dashed line shown in panels a and b. (d) Slice of the Hartree-Fock (red curve) and quasiparticle (green curve) STS images obtained along the horizontal dashed line shown in panels a and b.